\documentclass[pre,twocolumn,aps,showpacs]{revtex4-1}
\usepackage{graphicx}

\begin{document}
\title{Optical surface modes in the presence of nonlinearity and disorder
}
\author{M. I. Molina$^{1,2}$, N. Lazarides$^{3,4}$ and G. P. Tsironis$^{3,4}$
}
\affiliation{
$\ ^{1}$Departamento de F\'{\i}sica, Facultad de Ciencias, Universidad de Chile,
        Casilla 653, Santiago, Chile \\ 
$\ ^{2}$Center for Optics and Photonics, Casilla 4016, Concepcion, Chile \\
$\ ^{3}$Department of Physics, University of Crete, P.O. Box 2208,
        71003 Heraklion, Greece \\
$\ ^{4}$Institute of Electronic Structure and Laser,
        Foundation for Research and Technology-Hellas, P.O. Box 1527,
        71110 Heraklion, Greece
}

\date{\today}
\begin{abstract}
We investigate numerically the effect of the competition of disorder, 
nonlinearity, and boundaries on the Anderson localization of light waves
in finite-size, one-dimensional waveguide arrays. 
Using the discrete Anderson - nonlinear Schr\"odinger equation,
the propagation of the mode amplitudes up to some finite distance is monitored.
The analysis is based on the calculated localization length and the participation
number, two standard measures for the statistical description of Anderson 
localization. 
For relatively weak disorder and nonlinearity, a higher disorder 
strength is required to achieve the same degree of localization at the edge 
than in the interior of the array, in agreement with recent experimental
observations in the linear regime.
However, for relatively strong disorder and/or nonlinearity, this behavior is 
reversed and it is now easier to localize an excitation at the edge than in the 
interior.
\end{abstract}


\maketitle
{\em Introduction.-
}
A fundamental question concerning systems which are both disordered and nonlinear 
is whether or not Anderson localization \cite{Anderson1958} is weakened by the 
presence of nonlinearity. While it was originally developed in order to understand 
electronic transport in non-periodic (disordered) solids, the concept of 
Anderson localization was later generalized to the localization of classical 
waves in disorder media \cite{John1987}. 
The interaction of propagating
waves, and in particular of electromagnetic waves, when both disorder and 
nonlinearity are present can significantly affect localization and other 
phenomena \cite{Bishop1989}.

Despite of many efforts, that question has not been conclusively answered
\cite{Molina1998,Kopidakis2000,Kopidakis2008,MVK,Kivshar1990,Pikovsky2008,Flach2009,Fishman2011}.
It thus seems that the answer depends on the relative strength of disorder
and nonlinearity.  For large nonlinearity,
time-periodic and exponentially localized excitations in the form of discrete
breathers may be generated, due to the self-trapping effect \cite{Molina1993}. 
For small disorder strength, the discrete breathers are modulated to become
localized modes \cite{Ivanchenko2009}.
The above theoretical results were accompanied by a series of experimental
demonstrations of Anderson localization in optics \cite{Pertsch2004} and 
Bose-Einstein condensates \cite{Billy2008}.

It was recently observed experimentally that Anderson localization in finite
segments of
disordered waveguide arrays in the linear regime is actually site-dependent 
\cite{Szameit2010}.
Specifically, a higher disorder strength is required to achieve the
same degree of localization at the edge than in the interior (i.e., the "bulk")
of the array \cite{Szameit2010}.
Here we are interested in the effect of the interplay of disorder and nonlinearity
on the site-dependence of wavepacket localization in one-dimensional (1D),
disordered, finite-size arrays of coupled Kerr-type waveguides.
Using the discrete nonlinear Schr\"odinger (DNLS) equation with diagonal
(on-site) disorder,
frequently referred to as the discrete Anderson - nonlinear Schr\"odinger 
(DANLS) equation, 
we calculate standard measures of Anderson localization in order to analyze 
the site-dependence of the degree of localization for a wide range of
nonlinearity and disorder strengths.

{\em Model equations and statistical measures.-
}
Consider a 1D array of $N$ single-mode optical waveguides.  
In the framework of the coupled-modes formalism, the electric field $C(x,z)$	
propagating along the waveguides can be expanded as a superposition of the 
waveguide modes, $C(x,z) =\sum_{n}C_{n}(z)\phi(x-n)$, where $C_{n}$ is the 
complex amplitude  of the single guide mode $\phi(x)$ centered at the $n$th site. 
The evolution equations	for the modal amplitudes $C_{n}$ are
\begin{equation}
  \label{1}
   i\frac{d C_n}{dz} +V_{n,n-1} C_{n-1} 
          +\epsilon_n C_n +V_{n,n+1} C_{n+1} +\chi |C_n|^2 C_n =0 ,         
\end{equation}
where $n=1,2,3,...,N$, 
$\epsilon_n$ is the propagation constant associated with the $n$th site,
$V_{n,n\pm 1}$ are the tunneling rates between two adjacent sites,
$\chi$ is the nonlinearity parameter, and $z$ is the spatial coordinate
along the propagation direction (`time').
Eq. (\ref{1}) describes very well recent experiments in 1D
disordered waveguide lattices and, moreover, 
it serves as a paradigmatic model for a wide class of physical problems
where both disorder and nonlinearity are important.
Disorder is introduced into the optical lattice by randomly choosing  the 
propagation constants $\epsilon_n$ from a uniform, zero-mean distribution 
in the interval $[-\Delta, +\Delta]$.
As a result, the lattice remains periodic on average and, to a very good 
approximation, the parameters $V_{n,n\pm 1}$ become independent of the site 
number $n$, i.e., $V_{n,n\pm 1} =V$.
Then, Eq. (\ref{1}) reads
\begin{equation}
   i\frac{d C_n}{dz} +\epsilon_n C_n +V ( C_{n-1} +C_{n+1} )
         +\chi |C_n|^2 C_n =0 .  \label{2}
\end{equation}
In order to take into account the termination of the structure, 
we impose free boundary conditions at the edges, i.e., $C_0 = C_{N+1} =0$.
For $\chi \rightarrow 0$, Eq.(\ref{2}) reduces to the original Anderson model 
while in the absence of disorder ($\epsilon_{n}=0$), it reduces to the 1D 
DNLS equation \cite{Kevrekidis2009} that is generally non-integrable and it
conserves the norm ${\cal N}=\sum_{n=1}^N |C_n|^2$ and the Hamiltonian $\cal{H}$. 

To investigate the simultaneous interplay of disorder, nonlinearity and
boundary effects, we place initially a single-site excitation $C_n = \delta_{n,n_0}$
near, or at the boundary of the array.
This determines the value of the norm ${\cal N}=1$ for all subsequent `times'.
For a quantitative analysis we utilize two of the standard measures used 
in the description of Anderson localization; the participation number
$P = \left\{ \sum_{n=1}^N |C_n|^4 \right\}^{-1}$ , 	  
and the localization length $\ell$,
defined as the width of the envelope containing the localized profile.
The participation number gives a rough estimate of the number of sites where 
the wavepacket has significant amplitudes, and it is a useful measure for
ascertaining localization effects in the case of partial localization.
In this case, $P$ will saturate at a finite value, indicating the formation
of a localized wavepacket.

{\em Statistical analysis.-
}
In the following, we set $V=1$, while the nonlinearity  parameter $\chi$ 
varies between $0$ and $10$, and the disorder  width $\Delta$ takes on 
several different values. 
Since Anderson localization is essentially a statistical phenomenon, many
realizations of disorder are needed to obtain meaningful averages for the 
quantities of interest. This is particularly true for low-dimensional systems.
We typically use $n_R =1000$ realizations in each run. 
The array contains $N=200$  waveguides, and the maximum evolution ``time''
is $z=100$ (except otherwise stated).
In optics, nonlinearity is varied by changing the power content of the input beam.
However, this is formally equivalent to keeping the norm of the wavepacket 
fixed, and to varying the nonlinearity parameter $\chi$.
Eqs. (\ref{2}) are integrated with a standard 4rth order Runge-Kutta
algorithm with fixed time-stepping. 
We compute the absolute squared profiles $<|C_n|^2>$, where the brackets
denote averaging over all realizations $n_R$, hereafter referred to as 
Anderson mode profiles.
Assuming that the Anderson modes have a $z-$dependence which is a simple 
exponential function of the form $<|C_n|^2> = C_{max}^2 \, e^{-z/\ell}$,
the localization length $\ell$ can be computed via  $\chi^2$ fitting
procedure, with $C_{max}^2$ being the numerically obtained maximum of $<|C_n|^2>$.

\begin{figure}[t!]
\includegraphics[angle=0, width=0.85\linewidth]{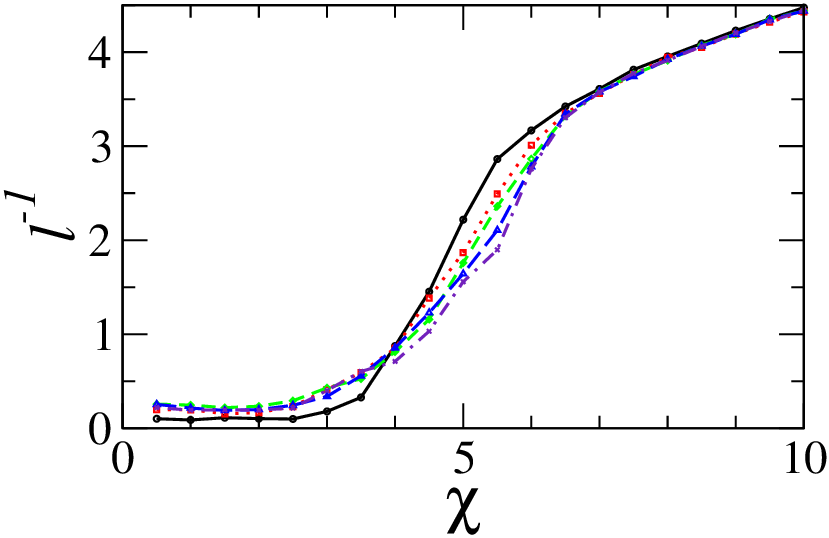}
\caption{(color online)
Inverse localization length $\ell^{-1}$ as a function of the nonlinearity 
strength $\chi$. 
A single-site excitation is launched from $n_0=1$ 
(solid - black), 2 (red - dotted), 3 (green - dashed), 5 (blue - long-dashed),
10 (indigo - dotted-dashed), 
for a system with $N=200$, $V=1$, $z=100$, $n_R=1000$, and $\Delta =0.6$. 
}
\end{figure}
\begin{figure}[t]
\includegraphics[angle=0, width=0.85\linewidth]{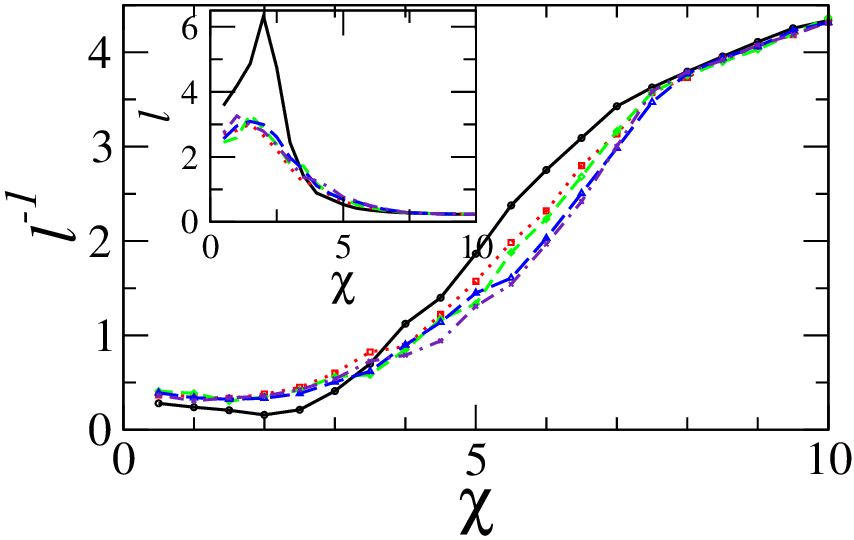}
\caption{(color online)
Inverse localization length $\ell^{-1}$ and localization length $\ell$ (inset)
as a function of the nonlinearity strength $\chi$.
Same parameters and notation as in Fig. 1, but $\Delta =1.0$.
}
\end{figure}
\begin{figure}[t]
\includegraphics[angle=0, width=0.85\linewidth]{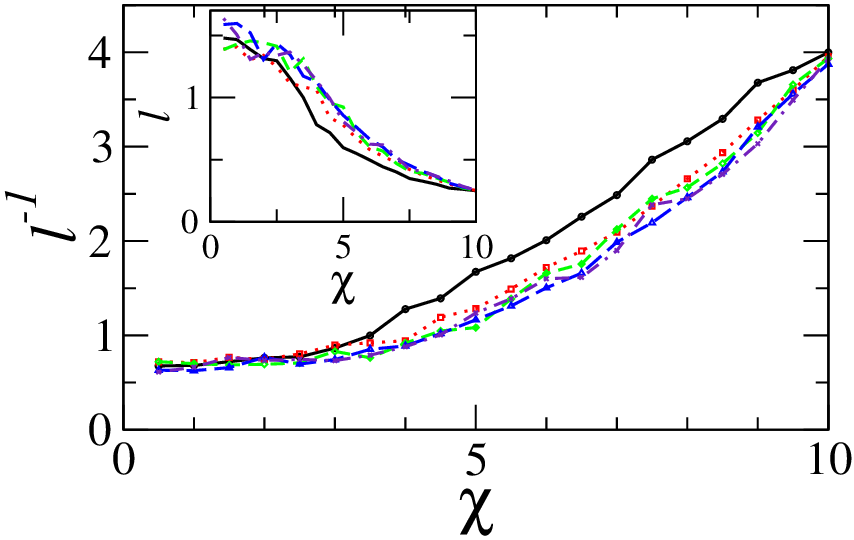}
\caption{(color online)
Inverse localization length $\ell^{-1}$ and localization length $\ell$ (inset)
as a function of the nonlinearity strength $\chi$.
Same parameters and notation as in Fig. 1, but $\Delta =2.0$.
}
\end{figure}

In Figs. 1-3 the inverse localization length $\ell^{-1}$ is shown as a function 
of $\chi$ for three different values of disorder strength.
In all the cases displayed in these figures the initial wavepacket is a single-site
excitation placed at $n=n_0$, with $n_0=1$ (right at the edge), 2, 3, 5, and 10. 
In Fig. 1 (where $\Delta=0.6$) we easily identify two different $\chi-$ regimes;
the weak and the strong nonlinearity regime, where $\ell^{-1}$ is small and large,
respectively. That generaly implies a lower degree of Anderson localization in 
the weak nonlinearity regime compared to that in the strong nonlinearity regime.
The large $\ell^{-1}$ in the interval of $\chi$ values where all the curves fall
the one onto the other, indicates the existence of a highly localized mode due 
to the self-trapping effect.
The characteristic nonlinearity strength, $\chi_c$, that roughly distinguishes between 
the two regimes is the critical on for self-trapping to occur in the 1D DNLS 
equation \cite{Molina1993}.
In the weak nonlinearity regime another important feature appears;
as it can be seen in the figure, the $\ell^{-1} (\chi)$ curve obtained for 
excitations initially placed at the edge ($n_0=1$) is well below all the others
(for which $n_{0}>1$).
Thus, an excitation initially placed at the edge ($n_0=1$) leads to final
wavepackets that are less localized than those which have been initialized 
below the 'surface' ($n_{0}>1$). 
This effect can be understood as the "repulsive" action of the boundary, 
reported in a previous work for surface modes in nonlinear periodic lattices
\cite{MVK}, and it is in agreement with the experimental observations of 
Ref. \cite{Szameit2010}. 

As the disorder strength $\Delta$ is increased from $0.6$ to $1.0$ (Fig. 2),
all the curves become flatter without showing any qualitative difference from 
those of Fig. 1. 
When $\Delta$ is increased to $2.0$, however, we do observe qualitative
differences (Fig. 3). For weak nonlinearity ($\chi \stackrel{<}{\sim} 4 \sim \chi_c$)
there are no significant differences in 
the degree of localization for the Anderson modes resulting from initial 
excitations either at the edge or in the interior of the array.
Thus, it is as easy to localize a wavepacket at the edge as it is in the
interior in this case. However, for intermediate nonlinearities
(from $\chi \stackrel{>}{\sim} 4$ to $\chi \simeq 8$) it is more favorable to loacalize
a wavepacket at the edge than in the bulk, whereas for large nonlinearities
$\chi \stackrel{>}{\sim} 8$ it is as easy to localize a wavepacket at the edge 
as it is in the bulk.
Thus, for relatively strong disorder we observe a bahavior that is strikingly 
different to what is observed in Figs. 1 and 2. 
The two different behaviors can be seen even more clearly by comparison of
the localization length $\ell$ as a function of $\chi$ for $\Delta=1$ and $2$,
shown in the insets of Fig. 2 and 3, repsectively. 
Thus, the presence of strong disorder is capable of overcoming the "repulsive"
character of the boundary for any value of $\chi$ and, moreover, it favors 
wavepacket localization at the edges for intermediate nonlinearities. 
\begin{figure}[t!]
\includegraphics[angle=0, width=0.85\linewidth]{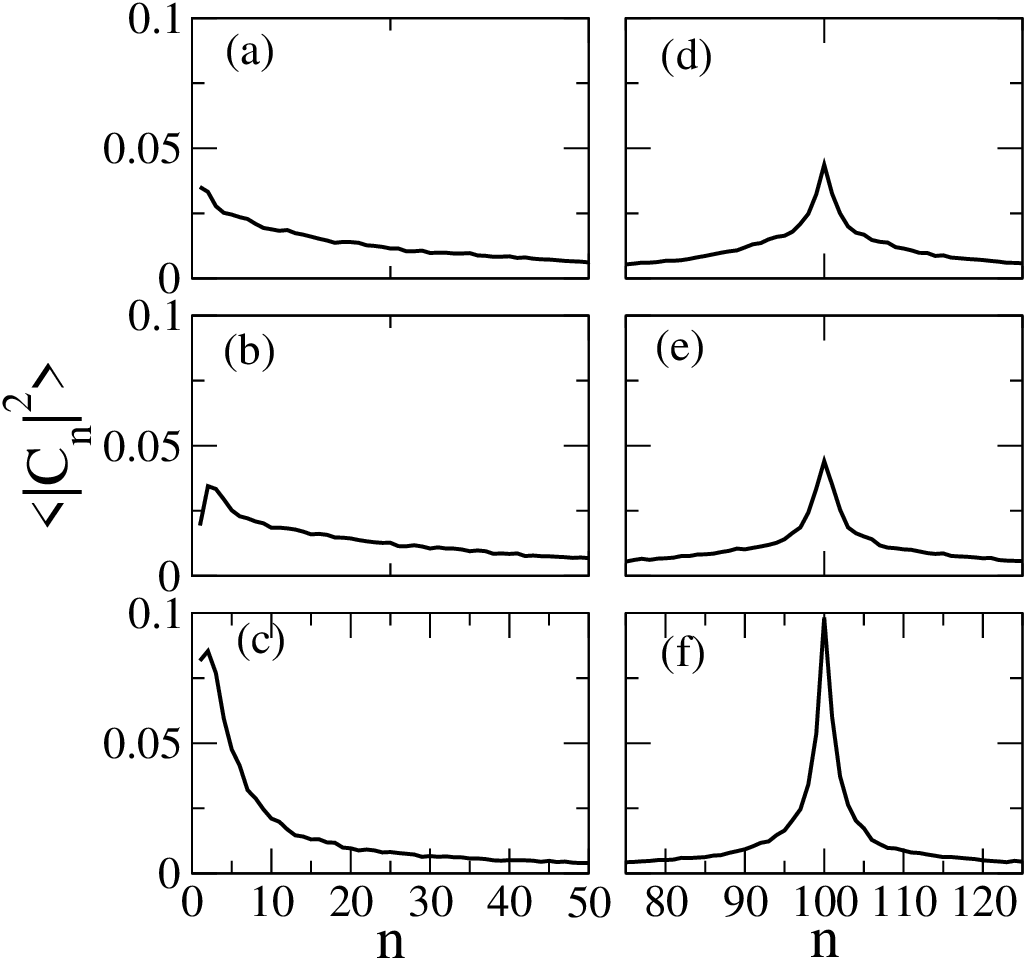}
\caption{Anderson mode profiles $<|C_n|^2>$ as a function of the
site-number $n$ 
for $N=200$, $V=1$, $\Delta=0.6$, $z=1000$, $n_R=1000$, 
and $\chi=1$ (a,d), $\chi=2$ (b,e) and $\chi=3$ (c,f).
Left panels denote the surface mode case ($n_{0}=1$) while right panels 
refer to the bulk mode case ($n_{0}=100$). 
Only part of the array sites are shown for clarity. }
\end{figure}

Typical examples of localized mode profiles both at, or close to the 
edge and the 'bulk' are shown in Fig. 4, where the Anderson mode
profiles $<|C_n|^2>$ are shown as a function of $n$ for $\Delta=0.6$.  
(Note that in this figure $z=1000$.)
The edge-localized modes are significantly more extended than the bulk modes,
even though the former are not all localized exactly at the edge. 
This is because of the small disorder strength $\Delta$, which
allows the repulsive force of the boundary on the mode to dominate
and move slightly the mode-maximum towards the bulk.
However, similar profiles (not shown) are obtained also for $\Delta=1.0$,
that is large enough to keep the localized modes at their initial location.
\begin{figure}[t!]
\includegraphics[angle=0, width=0.85\linewidth]{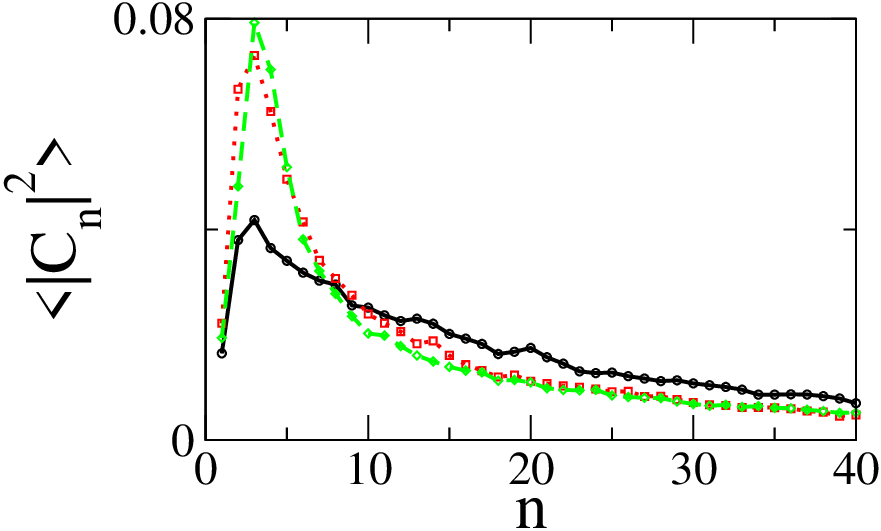}
\caption{(color online)
Averaged absolute squared profiles $<|C_n|^2>$ as a function of the 
site-number $n$ 
for $N=200$, $V=1$, $\Delta=0.4$, $n_R=1000$, $\chi=2.5$, $z=100$.
These profiles result from a single-site initial excitation at
$n_0 =1$ (black-solid); $n_0 =2$ (red-dotted); $n_0 =3$ (green-dashed).
Only part of the array sites are shown for clarity.
}
\end{figure}
 
Moreover, single-site excitations initialized at different sites $n_0$ 
can be pushed by the boundary towards the interior and form Anderson
modes at the same final site. 
These modes are different, at least for finite propagation distances $z$;
they differ in the degree of localization, leading to a multiplicity of 
Anderson modes having their maximum at the same site of the lattice (Fig. 5). 
For the particular value of $\chi$ used for Fig. 5, three single-site
excitations initialized at different $n_0$ have formed, 
after they have been propagated up to $z=100$, three distinct Anderson 
localized modes whose maximum is located at the same lattice site
(i.e., at $n=3$). 

\begin{figure}[t!]
\includegraphics[angle=0, width=0.85\linewidth]{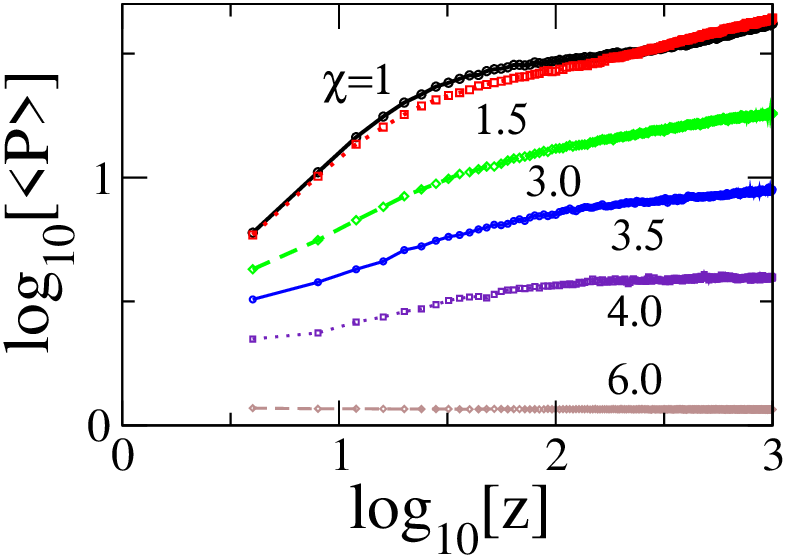}
\caption{(color online)
Logarithm of the averaged participation number $\log_{10}(<P>)$ 
as a function of $\log_{10}(z)$ 
for the surface case ($n_{0}=1$) and 
$N=200$, $V=1$, $\Delta=0.6$, $n_R=1000$, $z=1000$.
The values of $\chi$ are shown on the figure.
}
\end{figure}
\begin{figure}[t!]
\includegraphics[angle=0, width=0.85\linewidth]{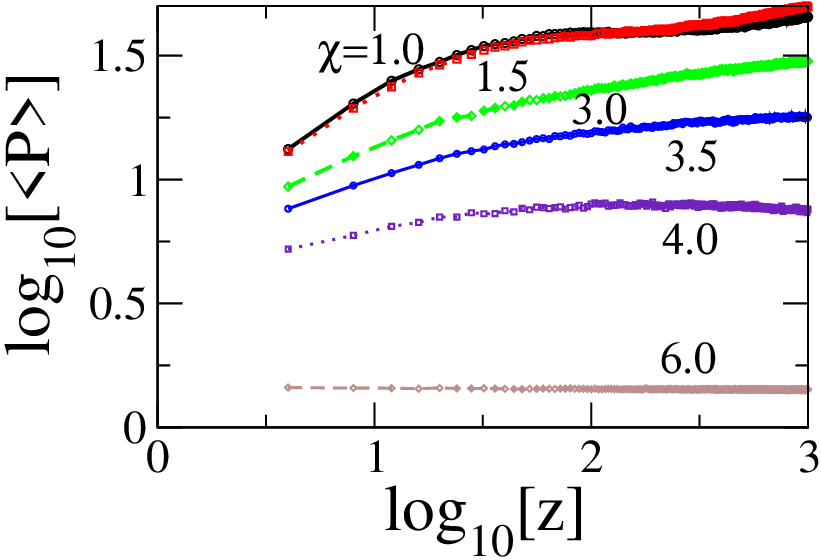}
\caption{(color online)
Logarithm of the averaged participation number $\log_{10}(<P>)$ 
as a function of $\log_{10} (z)$ for the bulk case ($n_{0}=100$) and  
$N=200$, $V=1$, $\Delta=0.6$, $n_R=1000$, $z=1000$, $n_0=100$.
The values of $\chi$ are shown on the figure.
}
\end{figure}
Finally, let us look at the participation number $P$ as a function of $z$ 
for wavepackets that are initially localized at the edge ($n_0=1$) and in the 
'bulk' ($n_0=100$) of the waveguide array. 
The logarithmic plots are shown in Fig. 6 and Fig. 7, respectively,
for several values of $\chi$ and weak disorder.
Comparing the curves in these figures corresponding to the same $\chi$,  
we see that those for $n_0=100$ are shifted to higher $P$ values 
than those for $n_0=1$. Thus, single-site excitations initialized at $n_0=100$ 
have, while propagating along $z$, a larger number of sites where the 
wavepacket has significant amplitudes. 
However, the excitations initialized at $n_0=100$ exhibit a higher degree
of localization than those initialized at $n_0=1$ (see also Fig. 4).
It is also interesting to see how the curves in each figure change as a 
function of $\chi$. For $\chi =6$, well above $\chi_c$, we see that the
wavepacket remains localized at a single site, independently of $n_0$.
For $\chi =4 \sim \chi_c$, the $\log_{10}(<P>)$ vs. $\log_{10} (z)$ curve
increases slowly with increasing $\log_{10} (z)$, and it saturates at a 
finite value around $z\simeq 100$, indicating the formation of a wavepacket
highly localized around $n_0$. For the nonlinearity strengths that are 
less than $\chi_c$, the $\log_{10}(<P>)$ vs. $\log_{10} (z)$ curves exhibit 
qualitatively the same behavior. There is an increase with increasing 
$\log_{10} (z)$ which is slowed down after some $z$ specific to each $\chi$ 
value, and indicates significant delocalization of the initially single-site
wavepacket. Delocalization is stronger for decreasing nonlinearity strength. 
However, those curves do not seem to saturate, implying that the corresponding
Anderson localized modes may delocalize further at longer propagation 
distances. 

{\em Concluding remarks.-
} 
We have performed extensive calculations with the 1D DANLS equation in order to
clarify some aspects of the interplay between boundary effects, disorder and 
nonlinearity, in finite-size waveguide arrays. 
In particular, we attempt to clarify the site-dependence of Anderson localization
that results from that interplay. We computed two standard measures of 
localization for discrete systems for varying nonlinearity and disorder strengths,
and we observed two strikingly different behaviors depending on the strength of 
the disorder.
For weak to moderate disorder, we distinguish two different nonlinearity regimes;
weak and strong, for values of $\chi$ roughly below and above $\chi_c$, 
respectively. 
In the weak nonlinearity regime it is easier to 
localize a wavepacket in the interior of the array than at the edge, which is 
in agreement with the experiments in the linear regime \cite{Szameit2010}.
In the strong nonlinearity regime it is as easy to localize a wavepacket at the 
edge as it is in the interior.
However, for relatively strong disorder, this behavior is reversed,
at least for intermediate nonlinearities, and it is now easier to localize
a wavepacket at the edge than in the bulk. 
For weak and very strong nonlinearities there is no significant site-dependence
on the degree of wavepacket localization.  
The results presented here obviously hold for finite propagation distance $z$, 
an important case of practical interest for experimentalists.

{\em Acknowledgements.-
} 
M.I.M. acknowledges support from Fondecyt Grant 1080374 and Programa de 
Financiamiento Basal de Conicyt (Grant No. FB0824/2008)

\end{document}